# WSN and Fog Computing Integration for Intelligent Data Processing

Viorel Mihai, Cristina Elena Hanganu, Grigore Stamatescu, Dan Popescu
Department of Automatic Control and Industrial Informatics
University Politehnica of Bucharest, 060042 Bucharest, Romania

*Abstract* – Networked embedded systems endowed with sensing, computing, control and communication capabilities allow the development of various application scenarios and represent the building blocks of the Internet of Things (IoT) paradigm. Traditional data collection methods include multiple field level IoT systems that can relay data stemming from a network of distributed ground sensors directly to a cloud platform for storage, analysis and processing. In such applications however, rapid sensor deployment in unstructured environments represents a challenge to the overall robustness of the system. We discuss the fog and mist computing approaches to hierarchically process data along its path from source to destination. The several stages of intermediate data processing reduce the computational and communication effort in a gradual manner. A three-layer topology for smart data monitoring and processing is thus proposed and illustrated to improve the information to noise ratio in a reference scenario.

## I. INTRODUCTION

In a world filled with digital innovations, technologies, including IoT, 5G Wireless Networking and embedded artificial intelligence (AI), are making rapid progress. These technologies are used to measure, monitor, process, analyze and react to a seemingly endless data deluge. One of the main recent advances in this area has been, the handling of the data by means of cloud computing. However these systems also become strained when physical world data is pushed in large quantities and fast rates. According to a forecast from the company Ericsson, the number of connected devices will reach 29 billion by 2022 [1]. Due to this increase both end-users and suppliers will have to face the scalability challenge. The need for a new approach is clear in terms of new architecture built to resolve latency, network inaccessibility, raising costs and sensitive data security concerns [2].

The paradigms of fog and mist computing are coming to help by proposing a new distributed architecture based on three levels for data collection and processing – Figure 1. Mist computing in this case is placed at the bottom of the hierarchy, near the field level devices: sensors and actuators, in the middle, fog computing handles and relays the data towards the cloud, which now represents the top layer. All three levels allow the implementation of critical core functions such as: compute, communication, control, storage and decision making. The advantage occurs when the data is processed at each level, by transmitting a much lower data volume towards the top level. Therefore, the response time decreases and resources such as computing power and the communication bandwidth will be managed in favor of large scale IoT systems. This architecture supports data quality at each level and improves the information/noise ratio. Thus, there are some guidelines principles [3]: relevant data is used at each level; communication between levels is done only when information is required; the network is dynamically configured and real-time; end devices must be aware of user needs and the local situation.

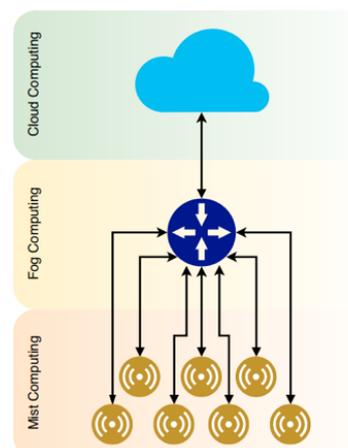

Figure 1. Mist – Fog – Cloud Computing hierachical topology

The rest of the article is structured as follows. Section II discusses related work in the context of the proposed hierarchical data processing architecture. In section III, the proposed algorithms and implementation steps are detailed. Section IV includes all the replicable experimental results achieved using a dedicated simulation environment on reference datasets. Section V concludes the paper by summarizing the main contributions.

## II. RELATED WORK

Cloud technology is based on Internet services delivery. Even though the idea is not new, it has impacted IT sector and helped a lot at IoT systems constraints. A definition provided by the National Institute of Standard and Technologies describes cloud computing as a ubiquitous model which provides resources e.g. servers, storage, computation power, offering a solution to the needs of IoT systems [3]. In the context of the emergence of this technology, large companies such as Amazon, IBM, Google, Facebook have created platforms to deliver cloud services as a business, for both economic and technical benefits [4].

In [5] cloud computing architecture is structured in four layers: hardware, infrastructure, platform and application. Cloud technology includes a centralized data collection and processing system which can provide a solution in many situations.

Despite of significant benefits of cloud computing, there are some issues when the number of devices directly connected increases substantially: energy management, high bandwidth consumption and computing power.

In a fog computing scenario, computation is performed at the gateway level. According to [6] fog computing is an extension of the cloud with a distributed architecture which comes to support several IoT applications such as: smart grid, smart cities, connected vehicles and wireless sensor networks (WSN). Hong et al. describes in [7] the fog computing programming models and simulation environments such as Mobile Fog and IFogSim. IFogSim proposes a simulation framework for IoT, Fog and Edge computing environments. It allows investigation of resource management techniques on the QoS criteria such as latency under different workloads.

If fog computing brings computation closer to the ground, at gateway and router level, mist computing advances this goal, closer to the field levelsensors and actuators. While the cloud and fog systems manage information globally, mist computing is locally aware of the application context. This level has the property of adapting, the configurations being realized at a sensor level whose functions are dynamic and adjustable [8]. The self-awareness of each device increases the autonomy of the network and brings important benefits such as reduced latency, long delays or bandwidth loading with redundant data [9].

### III. DATA PROCESSING ALGORITHM

The proposed architecture was implemented in iFogSim [10], a java simulation environment for Fog topologies. The framework was adapted to facilitate multi-layer data processing, as shown in Figure 2.

Firstly, the focus is on improving the ratio between useful information and noise. At sensor level, also named Mist computing layer basic procedures can be implemented to reduce the generated data volume. The data aggregation algorithm was developed as simple as possible, considering the low power and low memory requirements. The algorithm is described in [11], as a basic event triggering procedure. Basically, the measured data is send through gateway if the value exceeds an established bandwidth around the sliding average. The event E is defined as:

$$E = \{e(v_i) \in Q | T_1 > v_i > T_2\} \quad (1)$$

where:
- Q is the event set
- $v_i$ is the measured value at iteration i;
- $T_1$ and $T_2$ are the threshold defines as:

$$T_1 = as_i + p * as_i \quad (2)$$
$$T_2 = as_i - p * as_i \quad (3)$$

where:
- p is the bandwidth spawn.
- $as_i$ is the sliding average, defined as:

$$as_i = \frac{1}{n}\sum_{j=i-n}^{i-1} v_j \quad (4)$$

where:
- n is the range for $as_i$
- $v_j$ is the measured value at iteration j

The resulted mist computing algorithm is displayed in Table I. By comparing the physical value to the acceptance band defined around the sliding average, we can decide, at each measurement time stamp, if the value is relevant for the system.

TABLE I. MIST COMPUTING ALGORITHM

| Algorithm 1: Mist Computing Algorithm |
|---|
| Input: Raw measured data |
| Output: Processed data |
| Initialize:<br>  n = 10;  // sliding average domain<br>  p = 5%; // bandwidth spawn<br>**At each** Time-stamp<br>**SensorWakeUp**;<br>**read** sensorRawValue;<br>**if** E = true // the value is relevant for the system<br>    **Send** sensorRawValue;<br>**end if**<br>**Update** $as_i$<br>**Update** $T_1$ and $T_2$<br>**SensorToSleep**;<br>**end** |

Further on, the simulation environment iFogSim, is able to quantify the topology metrics in terms of bandwidth consumption and energy efficiency.

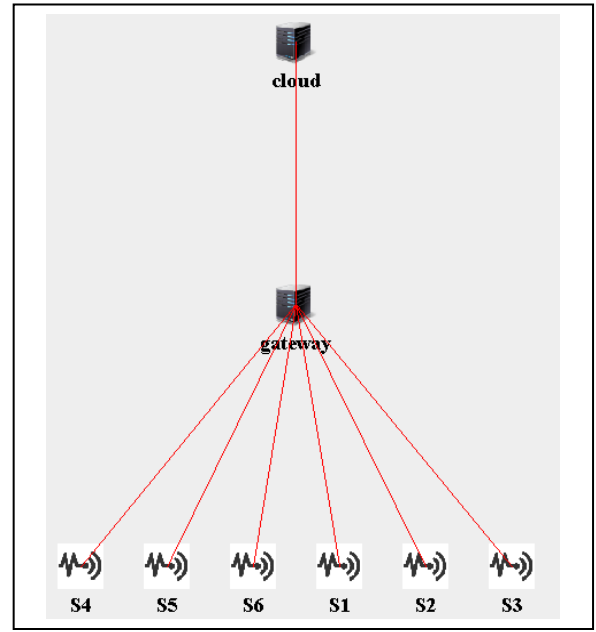

Figure 2. IFogSim Topology

The implementation was conducted as one sample network considering six IoT sensors, one fog computing gateway and one cloud platform. The topology structure was randomly selected. In table II are displayed the parameters for each topology component. Each sensor generates data determined by a normal distribution probabilistic model. The values are directed through the fog gateway in the cloud platform.

TABLE II.  IFOGSIM COMPONENT PROPERTIES

| Sensor | Properties |
|---|---|
| S1 | Normal distribution, mean value = 25, standard deviation = 4 |
| S2 | Normal distribution, mean value = 29, standard deviation = 8 |
| S3 | Normal distribution, mean value = 24, standard deviation = 2 |
| S4 | Normal distribution, mean value = 20, standard deviation = 6 |
| S5 | Normal distribution, mean value = 28, standard deviation = 1 |
| S6 | Normal distribution, mean value = 22, standard deviation = 6 |
| Gateway | Level = 1, uplink = 1000, downlink = 1000, Ram = 1000 |
| Cloud | Level = 0, uplink = 10000, downlink = 10000, Ram = 10000 |
| Link S1-gateway | Latency = 4 |
| Link S2-gateway | Latency = 6 |
| Link S3-gateway | Latency = 8 |
| Link S4-gateway | Latency = 2 |
| Link S5-gateway | Latency = 3 |
| Link S6-gateway | Latency = 7 |
| Link cloud- gateway | Latency = 50 |

IV. SIMULATION RESULTS

The experiments where conducted in two steps. First, the performances of the algorithm 1 where tested in a high-level programming environment. Online available raw data was used [12] as input for the algorithm. The raw dataset consists of 197853 indoor temperature datapoints, collected with a 1-minute time stamp. The considered time span is 20.07.2014 - 09.12.2014. The databased contains multiple sensors raw data streams measured in an office working space located in Richland, WA. For this experiment three such streams where selected.

TABLE III.  MIST COMPUTING ALGORITHM PERFORMANCES

| Sensor | Performances | | |
|---|---|---|---|
| | *Reduced data volume (data points)* | *Average error (°C)* | *Maximum error (°C)* |
| S1 | 196953 | 0.5983 | 2.7056 |
| S2 | 197597 | 0.3987 | 2.1833 |
| S3 | 190314 | 0.4536 | 2.2667 |

Three metrics where considered to appreciate the performance: the reduced data quantity and the average and maximum error between the input and output. In table III are presented the experiment results.

The results show that by accepting a measurement error, the topology can generate less data. It is worth to notice that the procedure can generate relatively high error in short therm. This can be controlled by fine tuning the sliding average domain (in the experiments n = 10) and the error bandwidth amplitude (in the experiments p = 0.1). The results are displayed in Fig.2-4.

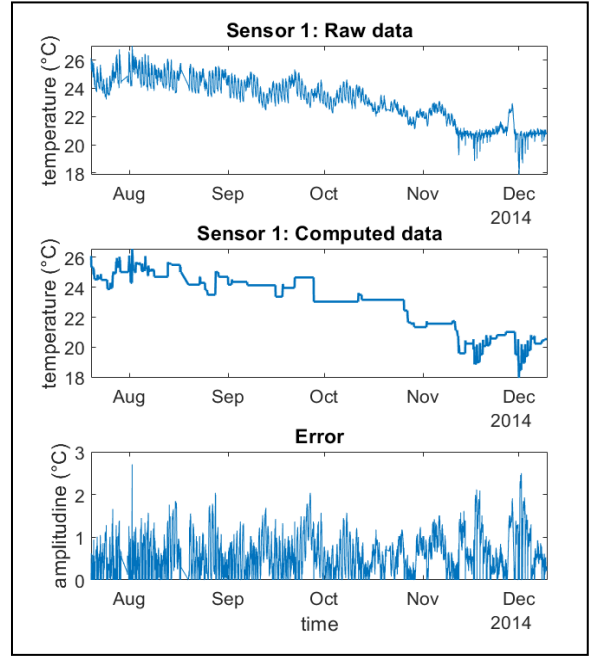

Figure 3.  Mist Computing algorithm performances for sensor 1

For the first sensor, an average error of approximately 2.6% was obtained. The generated data was reduced with 99.5 %.

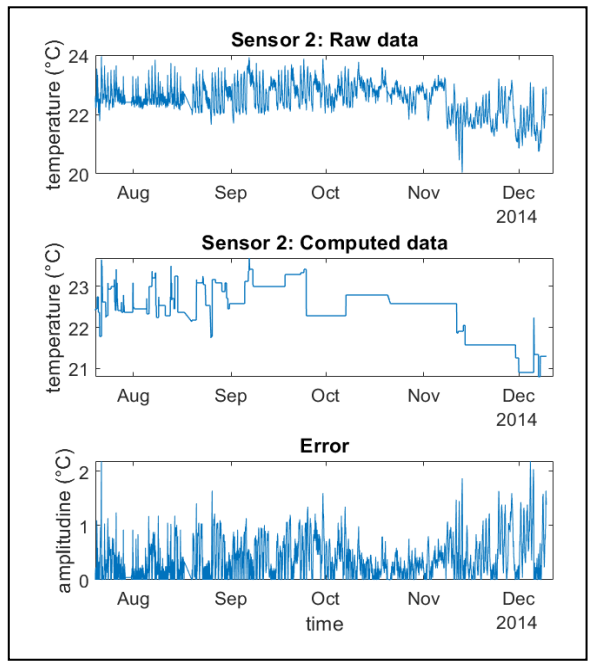

Figure 4.  Mist Computing algorithm performances for sensor 2

For the second sensor, an average error of approximately 1.8% was determined. The generated data was reduced with 99.8%.

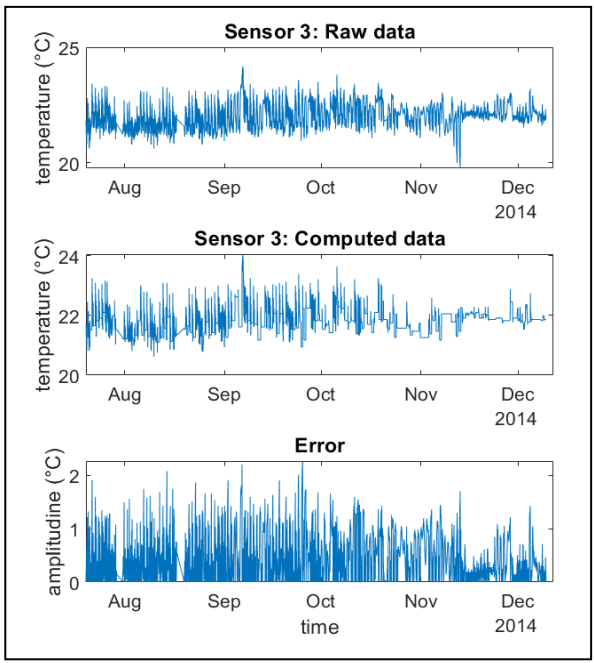

Figure 5. Mist Computing algorithm performances for sensor 3

For the third sensor, the average error is approximately 2%. The generated data was reduced with 96%.

For the second experiment, iFogSim was used to determine the network usage in usual cloud-based architecture vs. the mist-fog-cloud topology. The results show that in the second case, the network usage is reduced by 45.3%. The total energy consumption at cloud level was reduced with 0.5%.

## V. CONCLUSION

The article proposes an up-to-date three-layer topology: Mist-Fog-Cloud computing architecture. The objective was thus to reduce raw sensor data send towards cloud therefore the ratio between useful information and noise is improved. At the sensor level, basic data aggregation algorithm was implemented. As recent articles suggest, those algorithms are included in the mist computing layer. The article proposes an event triggered procedure that verify if the measured values is comprised between two threshold levels. Those thresholds are defined as an acceptance band around the sliding average. The full architecture was implemented in a simulator environment, named iFogSim. The implementation results consist basically in a network usage quantification.

As future work, the architecture will be implemented in a realistic scenario. An open topology will be contoured around the proposed architecture. The final configuration will follow the open hardware and open software philosophy mainly because for training and teaching reasons. At sensor level, Arduino-type MCU with suitable communication modules could serve as the hardware platform for the mist computing layer. At fog level, more powerful embedded Linux boards can be a suitable candidate for a local gateway implementation that facilitates the data streaming to the cloud.

ACKNOWLEDGMENT

This work has been partially funded through the project "Integrated Intelligent Monitoring System UAV-WSN-IoT for Precision Agriculture" (MUWI). Contract no. 1224/22.01.2018 - NETIO 53/05.09.2016.